\documentclass[aps,prl,twocolumn,epsfig,groupedaddress]{revtex4}

\usepackage{epsfig}
\begin{document}

\title{A Dynamical Scaling Law for Jet Tomography}

\author{Carlos A. Salgado and Urs Achim Wiedemann}
\address{Theory Division, CERN, CH-1211 Geneva 23, Switzerland}

\date{\today}

\begin{abstract}
Medium modifications of parton fragmentation provide a novel
tomographic tool for the study of the hot and dense matter
created in ultrarelativistic nucleus-nucleus collisions. Their
quantitative analysis, however, is complicated by the strong
dynamical expansion of the collision region. Here, we establish
for the multiple scattering induced gluon radiation spectrum
a scaling law which relates medium effects in a collision of
arbitrary dynamical expansion to that in an equivalent static
scenario. Based on this scaling, we calculate for typical
kinematical values of the RHIC and LHC heavy ion programs
medium-modified parton fragmentation functions for heavy
ion collisions with realistic dynamical expansion.
\end{abstract}
\maketitle
 \vskip 0.3cm

For the study of the shortlived state of hot and dense matter
produced in nucleus-nucleus collisions, novel tools become
available at collider energies. In particular, the abundant
production of high-$p_\perp$ partons in hard processes dominates
the high-$p_\perp$ tails of hadronic particle spectra.
These partons propagate through the medium while fragmenting into
hadrons. Basic properties of the hot and dense matter such as
the average in-medium pathlength of hard partons and the
transverse colour field strength (energy density) they experience
are thus reflected in the medium-dependence of parton
fragmentation.

Motivated by this idea~\cite{Gyulassy:1993hr}, several
groups~\cite{Baier:1996sk,Zakharov:1997uu,Wiedemann:2000za,Gyulassy:2000er}
calculated in recent years gluon radiation spectra due to
medium-induced multiple scattering contributions in order to
understand the medium-dependence of hadronic cross sections.
The aim of this letter is to extend previous
studies~\cite{Baier:1998yf,Gyulassy:2000gk} to the full gluon
radiation spectrum in
a strongly expanding medium of small finite size, to
calculate for the first time the medium-dependent fragmentation
functions resulting from many soft interactions of the hard parton with
the medium, and to explore observable consequences for collider
experiments at RHIC and LHC.

Hadronic cross sections for high $p_\perp$ particle production
are calculated by convoluting the parton
distributions of the incoming projectiles with the product $d\sigma^h(z,Q^2)$
of a perturbatively calculable partonic cross section $\sigma^q$ and
the fragmentation function $D_{h/q}(x,Q^2)$ of the produced parton,
$d\sigma^h(z,Q^2) = \left( \frac{d\sigma^q}{dy}\right) dy\,
D_{h/q}(x,Q^2)\, dx$. Here, $x=E_h/E_q$, $y=E_q/Q$ and $z=E_h/Q$
denote fractions between the virtuality of the hard process $Q$,
and the energies of the produced parton and resulting hadron.
If the produced parton looses with probability $P(\epsilon)$
an additional fraction $\epsilon = \frac{\Delta E}{E_q}$  of
its energy due to medium-induced radiation, then the hadronic
cross section is given in terms of the medium-modified
fragmentation function~\cite{Wang:1996yh,Gyulassy:2001nm}
\begin{eqnarray}
  D_{h/q}^{(\rm med)}(x,Q^2) = \int_0^1 d\epsilon\, P(\epsilon)\,
  \frac{1}{1-\epsilon}\, D_{h/q}(\frac{x}{1-\epsilon},Q^2)\, .
  \label{eq1}
\end{eqnarray}
The probability that a hard parton looses $\Delta E$ of its
initial energy due to the independent emission of an arbitrary
number of $n$ gluons is determined by the medium-induced gluon energy
spectrum $\frac{dI}{d\omega}$,~\cite{Baier:2001yt}
\begin{eqnarray}
  P(\Delta E) &=& \sum_{n=0}^\infty \frac{1}{n!}
  \left[ \prod_{i=1}^n \int d\omega_i \frac{dI(\omega_i)}{d\omega}
    \right]
    \nonumber \\
    && \times
    \delta\left(\Delta E - \sum_{i=1}^n \omega_i\right)
    \exp\left[ - \int d\omega \frac{dI}{d\omega}\right]\, .
   \label{eq2}
\end{eqnarray}
We evaluate the quenching weight (\ref{eq2}) via its Mellin
transform (see~\cite{Baier:2001yt} for details), starting from the
medium-induced BDMPS-Z gluon radiation
spectrum~\cite{Wiedemann:2000za,Wiedemann:2000tf},
\begin{eqnarray}
  && \omega\frac{dI}{d\omega}
  = {\alpha_s\,  C_F\over (2\pi)^2\, \omega^2}\,
    2{\rm Re} \int_{\xi_0}^{\infty}\hspace{-0.3cm} dy_l
  \int_{y_l}^{\infty} \hspace{-0.3cm} d\bar{y}_l\,
   \int d^2{\bf u}\,  d^2{\bf k}\, e^{-i{\bf k}_\perp\cdot{\bf u}}   \,
  \nonumber \\
  && \times
  e^{ -\frac{1}{2} \int_{\bar{y}_l}^{\infty} d\xi\, n(\xi)\,
    \sigma({\bf u}) }\,
  {\partial \over \partial {\bf y}}\cdot
  {\partial \over \partial {\bf u}}\,
  {\cal K}({\bf y}=0,y_l; {\bf u},\bar{y}_l|\omega) \, ,
    \label{eq3}
\end{eqnarray}
where
\begin{eqnarray}
 {\cal K}
 = \int {\cal D}{\bf r}
   \exp\left[ i \int_{y_l}^{\bar{y}_l} d\xi
        \frac{\omega}{2} \left(\dot{\bf r}^2
          - \frac{n(\xi) \sigma\left({\bf r}\right)}{i\, \omega} \right)
                      \right]\, .
  \label{eq4}
\end{eqnarray}
Medium properties enter $dI/d\omega$ via the product of the medium
density $n(\xi)$ of scattering centers times the relative strength
of the single elastic scattering cross section
$\propto |a_0({\bf q}_\perp)|^2$ measured by the
dipole cross section
$\sigma({\bf r}) = 2\, C_A\, \int
  \frac{d{\bf q}_\perp}{(2\pi)^2}\, |a_0({\bf q}_\perp)|^2\,
  \left( 1 - e^{-i{\bf q}_\perp\cdot {\bf r}}\right)$.
We consider the effect of arbitrarily many soft momentum transfers to
the parton, when the path integral in (\ref{eq3}) can be
evaluated in the saddle point
approximation~\cite{Zakharov:1997uu,Wiedemann:2000za},
$\sigma({\bf r}) \approx C\, {\bf r}^2$. This is complementary
to studies which consider one additional medium-induced gluon
exchange via twist-4 matrix elements~\cite{Guo:2000nz,Qiu:2001hj}, or
up to $N \leq 3$ medium-induced momentum transfers in an opacity
expansion~\cite{Wiedemann:2000za,Gyulassy:2000er} of (\ref{eq3}).

For the case of a nuclear medium without dynamical evolution,
$n(\xi) = n_0$, the radiation spectrum (\ref{eq3}) depends
on the partonic in-medium pathlength $L$ and the transport
coefficient $n_0C$. The latter measures the squared average
momentum picked up by the partonic projectile per unit
pathlength~\cite{Wiedemann:2000tf}. Phenomenological
estimates~\cite{Baier:1996sk,Wiedemann:2000tf,Arleo} range typically
between $n_0C = (50\, {\rm MeV})^2/{\rm fm}$ for normal
nuclear matter and $n_0C = (500\, {\rm MeV})^2/{\rm fm}$
for a hot quark-gluon plasma.
Writing all energies in units of the characteristic gluon frequency
$\omega_c = 2\, n_0C\, L^2$, the probability $P(\Delta E/\omega_c)$
depends on only one further, dimensionless parameter combination
\begin{equation}
  R=n_0C\, L^3\, .
  \label{eq5}
\end{equation}
%
We refer to $R$ as ``density parameter'' since it characterizes
the initially produced gluon rapidity density (see eq. (\ref{eq12})  below).

Fig.~\ref{fig1} shows the quenching weights $P(\Delta E/\omega_c)$
calculated for a static nuclear medium and varying density parameters
$R$. In comparison to earlier studies~\cite{Baier:2001yt}, we observe
as a novel feature the occurence of a discrete contribution $p_0$ in
\begin{eqnarray}
  P({\Delta E/ \omega_c}) &=& p_0\, \delta({\Delta E/ \omega_c}) 
+ p({\Delta E/ \omega_c})\, .
   \label{eq6}
\end{eqnarray}
This is a consequence of considering a medium
of realistic small size and opacity, where the projectile
escapes the collision region with finite probability $p_0$
without further interaction. With increasing density parameter $R$,
the probability $p_0$ of no interaction tends to zero, and the width
of $p({\Delta E/ \omega_c})$ broadens since a larger energy fraction
$\Delta E$ is lost. In ~\cite{Baier:2001yt},
an analytical approximation for $P(\epsilon)$ is given solely
for illustrative purposes. It is based on the small-$\omega$
approximation $\frac{dI}{d\omega} \propto \alpha_s
\sqrt{\frac{\omega_c}{\omega}}$ of an expression valid for 
large in-medium pathlengths only. 
We find that it captures well the shape of $P$ for large systems
where $p_0 \ll 1$. However it shows an unphysical
large-$\epsilon$ tail with
infinite first moment $\int d\epsilon\, \epsilon\, P(\epsilon)$.
As seen from eq. (\ref{eq11}) discussed below, the accuracy of 
our study profits significantly
from a better numerical control of this large $\epsilon$ region. 
%
\begin{figure}[h]\epsfxsize=8.7cm
\epsfysize=11.7cm
\vspace{-1cm}
\centerline{\epsfbox{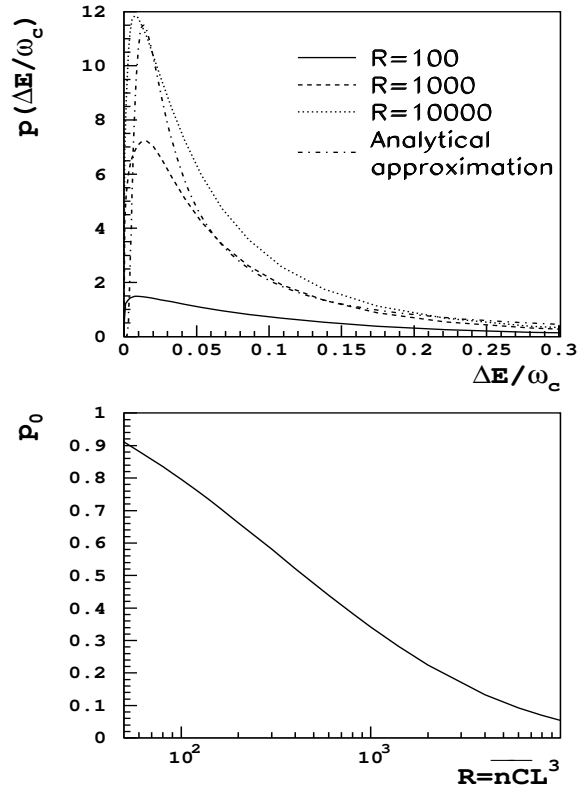}}
\vspace{-0.5cm}
\caption{
The two contributions to 
the probability (\ref{eq2}) that a parton looses $\Delta E$
of its energy in the medium: Continuous part 
(upper figure) and the discrete probability $p_0$ in (\ref{eq6}) 
that the hard parton
escapes the medium without interaction (lower figure).
}\label{fig1}
\end{figure}

The strong longitudinal and - to a lesser extent - transverse expansion
of a heavy ion collision implies a decreasing density of scattering
centers $n(\xi)$,
\begin{equation}
  n(\xi)C = n_dC \left( \frac{\xi_0}{\xi} \right)^\alpha\, .
  \label{eq7}
\end{equation}
Here, $\alpha = 0$ characterizes a static medium
and $\alpha =1$ corresponds to a one-dimensional, boost-invariant
longitudinal expansion consistent e.g. with hydrodynamical
simulations. The maximal value $n_d$ is reached around the 
formation time $\xi_0$ of the medium which can be set by
the inverse of the saturation scale $p_{\rm sat}$~\cite{EKRT},
resulting in $\approx$ 0.2 fm/c at RHIC and $\approx$ 0.1 fm/c at LHC. 
Since the difference between $\xi_0$ and the production time of the 
hard parton is irrelevant for the evaluation of the radiation spectrum 
(\ref{eq3}), we replace the latter in (\ref{eq3}) by $\xi_0$.

\begin{figure}[h]\epsfxsize=9.2cm
\centerline{\epsfbox{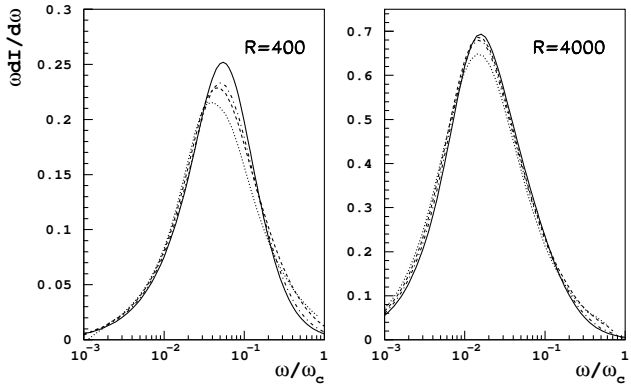}}
\caption{
The medium-induced gluon radiation spectrum
$\omega\, \frac{dI}{d\omega}$ for different dynamical expansion parameters
$\alpha$=0 (solid line), $\alpha$=0.5 (dotted-dashed), 
$\alpha$=1 (dashed) and $\alpha$=1.5 (dotted).
The scaling law (\ref{eq8}) is shown for R=400 and R=4000.
}\label{fig2}
\end{figure}

We evaluate the gluon radiation spectrum for dynamically
expanding collision regions over a wide range of expansion
parameters $\alpha\in [0:1.5]$ by combining the analytic
solution~\cite{Baier:1998yf} of the path-integral (\ref{eq4})
for a time-dependent harmonic oscillator with the treatment of
finite size effects~\cite{Wiedemann:2000tf}. We observe a simple
scaling law which relates the gluon radiation spectrum (\ref{eq3})
of a dynamically expanding collision region to an equivalent static
scenario. The linearly weighed line integral
\begin{equation}
  \overline{nC} = \frac{2}{L^2}\int_{\xi_0}^{\xi_0+L}
  d\xi\, \left(\xi - \xi_0\right)\,
  n(\xi)\, C
  \label{eq8}
\end{equation}
defines the transport coefficient of the equivalent static
scenario. Fig.~\ref{fig2} illustrates the accuracy of this
scaling law for expansion parameters $\alpha = 0$, 0.5, 1.0, 1.5,
and different values of the density parameter $R = \overline{nC} L^3$.
The accuracy improved with increasing density parameter $R$.
On the level of the quenching weight (\ref{eq2}), it
was better than 10 $\%$ in the physically relevant
parameter range, except for very thin media ($R < 100$) where
energy loss effects are negligible ($< 2 \%$). In a subsequent
publication~\cite{SW02b}, we shall extend this
scaling law to the ${\bf k}_\perp$-differential
radiation spectrum, and we shall document a CPU-inexpensive numerical
routine which allows to implement the quenching weight (\ref{eq2})
for arbitrary values of $\alpha$, $n_dC$ and $L$ in event generator
studies of hadronic spectra.

%
The linear weight in (\ref{eq8}) implies that scattering centers
which are further separated from the partonic production point $\xi_0$
are more effective in generating gluon radiation.
For a dynamical expansion following Bjorken scaling
[$\alpha = 1$ in eq. (\ref{eq7})], all timescales contribute
equally to $\overline{nC}$. This indicates that observables
related to medium-induced gluon radiation give access to
the quark-gluon plasma lifetime.

Since the radiation spectrum $\frac{dI}{d\omega}$ depends
on the expansion parameter $\alpha$, the length
$L$ and the transport coefficient only through $\overline{nC}$,
dynamical scaling holds automatically for the average parton 
energy loss
\begin{equation}
  \langle \Delta E \rangle \equiv \int dE\, E\, P(E)
  = \int d\omega\, \omega\, \frac{dI}{d\omega}\, .
  \label{eq9}
\end{equation}
This is seen already from the expressions for $\langle\Delta E\rangle$ 
derived for dynamically expanding scenarios by Baier et 
al.~\cite{Baier:1998yf} in the dipole approximation and by Gyulassy et
al.~\cite{Gyulassy:2000gk} for $N=1$ scattering center.
Moreover, Gyulassy et al.~\cite{Gyulassy:2000gk} conjectured that
higher order opacity terms ($N=2,3$) of $\langle\Delta E\rangle$
show the same dependence on $\overline{nC}$.
The results presented here go beyond establishing this conjecture and
confirming the result of Ref.~\cite{Baier:1998yf}. Our novel finding is
that dynamical scaling holds beyond the $\omega$-integrated average energy
loss (\ref{eq9}) to good numerical accuracy for the $\omega$-differential
gluon radiation spectrum. This is important since a reliable
calculation of the medium-dependence of hadronic spectra
requires~\cite{Baier:2001yt} knowledge of the quenching weight (\ref{eq2})
beyond the first moment (\ref{eq9}) [i.e., requires knowledge about
$\frac{dI}{d\omega}$]. Moreover, the scaling law established here
is relevant for the angular gluon radiation pattern~\cite{SW02b},
since the radiation spectrum emitted outside the opening angle $\Theta$
can be calculated in the present approach essentially by
replacing $R= n_0CL^3 \to \sin^2\Theta\, n_0CL^3$.

\begin{figure}[h]\epsfxsize=8.7cm
\vspace{-1.0cm}
\centerline{\epsfbox{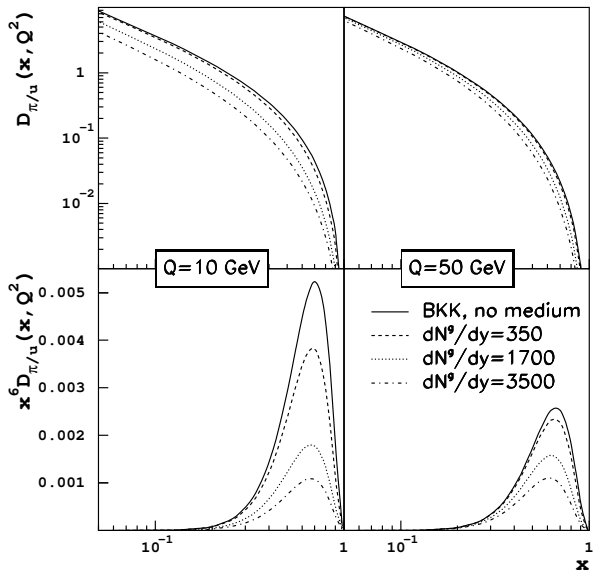}}
\caption{The LO BKK~\cite{Binnewies:1994ju} fragmentation function
 $u\to \pi$ for no medium and the medium-modified fragmentation
functions for different gluon rapidity densities (see eq. (\ref{eq12}))
and $L=7$ fm. 
}\label{fig3}
\end{figure}

We calculate the medium dependence of parton fragmentation
functions $D_{h/q}(x,Q^2)$ from (\ref{eq1}) using the quenching
weights (\ref{eq2}) and the LO BKK~\cite{Binnewies:1994ju}
parametrization of $D_{h/q}(x,Q^2)$. Typical results for
the pion fragmentation function of up quarks are shown
in Fig.~\ref{fig3}. For the present
study, we identify the virtuality $Q$ of $D_{h/q}(x,Q^2)$ with
the (transverse) initial energy $E_q$ of the parton.
This is justified since $E_q$ and $Q$ are of the same order, and
$D_{h/q}(x,Q^2)$ has a weak logarithmic $Q$-dependence while
medium-induced effects change as a function of
$\epsilon = \frac{\Delta E}{Q} \approx O(\frac{1}{Q})$.

As seen in Fig.~\ref{fig3}, the fragmentation function decreases for
increasing values of the transport coefficient $\overline{nC}$,
since the probability of a parton of initial energy $E_q$ to
fragment into a hadron of large energy $xE_q$ decreases with
increasing parton energy loss. The relative size of this medium
modification changes roughly like
$\epsilon = \frac{\Delta E}{Q} \approx O(\frac{1}{Q})$. We note
that the calculation based on (\ref{eq1}) is not reliable for small
fractions $x$ ($x < 0.1$ say), since it implies for
$D_{h/q}^{(\rm med)}(x,Q^2)$ a normalization
\begin{equation}
  \int_0^1 dx\, x\, D_{h/q}^{(\rm med)}(x)
  \simeq \int_0^1 dx\, x\, D_{h/q}(x)
    \int d\epsilon\, (1-\epsilon)\, P(\epsilon)\, ,
    \label{eq11}
\end{equation}
which is a factor $\int d\epsilon\, \epsilon\, P(\epsilon)$
too small. The origin of this error is that the hadronized remnants
of the medium-induced soft radiation are neglected in the definition
of (\ref{eq1}). Due to the softness of these remnants, however,
the true medium modified fragmentation function is underestimated
by equation (\ref{eq1}) for small $x$ only. We expect eq. (\ref{eq1})
to be valid for the calculation of $D_{h/q}^{(\rm med)}(x,Q^2)$ for
$x > 0.1$.

To illustrate
the effect of medium-modified fragmentation functions on hadronic
cross sections, we plot in Fig.~\ref{fig3}
the fragmentation function multiplied by $x^6$. This exploits
that hadronic cross sections weigh the fragmentation function
$D_{h/q}^{(\rm med)}(x,Q^2)$ by the initial hard partonic cross sections
${d\sigma^q}/{dp_{\perp}^2} \sim {1}/{p_{\perp}^{n(\sqrt{s})}}$
and thus effectively test $x^{n(\sqrt{s})} D_{h/q}^{(\rm med)}(x,Q^2)$.
We choose $n=6$ for illustrative purposes. For interpretation, the
position of the maximum $x_{\rm max}$ of  $x^6 D_{h/q}^{(\rm med)}(x,Q^2)$
corresponds to the most likely energy fraction $x_{\rm max} E_q$
of the leading hadron. The medium-induced relative reduction of
$x^6 D_{h/q}^{(\rm med)}(x,Q^2)$ around its maximum translates into
a corresponding relative suppression of this contribution to the
high-$p_\perp$ hadronic spectrum at $p_\perp \sim x_{\rm max} E_q$.
For $Q=10$ GeV, e.g., the leading hadron has most likely $p_\perp \sim$
5 - 8 GeV, and a reduction of this contribution to the pion spectrum by 
a factor $2$ corresponds e.g. to $L = 7$ fm and  
$\overline{nC} = $ 1 - 5 fm$^{-3}$, see Fig.~\ref{fig3}.
%
%
The transport coefficient can be related to the initial gluon 
rapidity density~\cite{Baier:1996sk,Gyulassy:2000gk}.  
Uncertainties in this procedure remain to be discussed. For 
the purpose of comparing our parameter values directly to 
results in the recent literature~\cite{Gyulassy:2000gk},
we use for a Bjorken scaling scenario 
%
%
\begin{equation}
  R = \overline{nC}L^3 = \frac{L^2}{R_A^2}\, \frac{dN^g}{dy}\, ,
  \quad \hbox{for}\quad \alpha =1\, ,
\label{eq12}
\end{equation}
where $R_A$ denotes the nuclear radius.
The curves in Fig.~\ref{fig3} thus correspond to 
$\overline{nC} \approx $ 1, 5 and 10 fm$^{-3}$, respectively.

A quantitative study of hadronic cross sections and a careful
discussion of theoretical uncertainties is clearly needed for a 
detailed comparison of the above results to measured hadronic 
spectra. It is nevertheless interesting to contrast Fig.~\ref{fig3}
with the transverse $\pi^0$-spectrum measured at RHIC~\cite{David:2001gk}. 
Taking the above-mentioned values at face value, the soft multiple scattering
approach to parton fragmentation functions relates a medium-induced
suppression factor $\approx 2$ at $p_\perp \approx 5$ GeV to an initial
gluon rapidity density which seems comparable with the value 500-1000
extracted in Ref.~\cite{Gyulassy:2000gk}. Thus, while the multiple soft
BDMS scattering approach without finite size treatment was
criticised~\cite{Gyulassy:2000gk} for overestimating the effects
of parton energy loss below $p_\perp \approx 10$ GeV by up to an order
of magnitude, the present findings and those
obtained by a lowest order opacity expansion seem comparable.
Differences pointed out~\cite{Gyulassy:2000gk} earlier do not allow
to prefer the $N=1$ opacity expansion over the multiple soft scattering
approach but just indicate that finite size effects have to be treated
in both approaches with the same scrutiny.

Equation (\ref{eq12}) suggests that the average transport coefficient
$\overline{nC}$ changes from RHIC to LHC proportional to the observed
particle multiplicity. Assuming a factor $\sim 5$ increase consistent
with recent studies~\cite{EKRT}, Fig.~\ref{fig3} suggests
the existence of a sweet spot at LHC energies in the $p_\perp$-range
around 50 GeV, where theoretical uncertainties are much better controlled
than at $p_\perp \leq 10$ GeV while suppression factors are still
sufficiently large (of order 2) to be experimentally accessible.
Attaching more precise numbers to this discussion lies beyond the
exploratory calculation presented here. It does not only require 
a quantitative understanding of hadronic spectra in ultrarelativistic
nucleus-nucleus collisions based on the knowledge of the underlying
partonic cross sections and the nuclear parton distribution functions. It
also requires medium-modified fragmentation functions for a realistic
distribution of in-medium pathlengths $L$ and initial densities. This
in turn requires modeling of the spatio-temporal evolution of the hot 
and dense medium. The scaling law established here simplifies the 
inclusion of these dynamical effects in quantitative studies. It can 
be extended to the study of the medium-dependence of angular radiation 
patterns~\cite{SW02b}.

We thank F. Arleo, K. Eskola, A. Kovner, K. Redlich 
and V. Ruuskanen for helpful discussions.
C.A.S. is supported by a Marie Curie Fellowship no.
HPMF-CT-2000-01025 of the European Community TMR program.



\end{document}